\documentclass[prl,a4paper,twocolumn,preprintnumbers,nofootinbib]{revtex4}

\usepackage[pdftex]{graphicx}
\usepackage{amsmath,amssymb}
\usepackage{mathrsfs,wasysym}
\usepackage[pdftex]{hyperref}

\newcommand{\mt}{m_{\rm t}}
\newcommand{\mh}{m_H}
\newcommand{\Lum}{\mathscr{L}}
\newcommand{\as}{\alpha_s}
\newcommand{\Ord}{\mathcal{O}}
\def\beq{\begin{equation}}  
\def\eeq{\end{equation}}
\def\({\left(}
\def\){\right)}
\def\[{\left[}
\def\]{\right]}
\allowdisplaybreaks

\begin{document}

\title{The threshold region for Higgs production in gluon fusion}
\author{Marco Bonvini}
\affiliation{Deutsches Elektronen-Synchroton, DESY, Notkestra{\ss}e
  85, D-22603 Hamburg, Germany}
\author{Stefano Forte} \affiliation{Dipartimento di Fisica,
  Universit\`a di Milano and INFN, Sezione di Milano, Via Celoria 16,
  I-20133 Milano, Italy}
\author{Giovanni Ridolfi} \affiliation{Dipartimento di Fisica,
  Universit\`a di Genova and INFN, Sezione di Genova, Via Dodecaneso
  33, I-16146 Genova, Italy}
\preprint{DESY 12-057}
\preprint{IFUM-994-FT}

\begin{abstract}
We provide a quantitative determination of the effective partonic
kinematics for Higgs production in gluon fusion in
terms of the collider energy at the LHC. We use the result to assess,
as a function of the Higgs mass, whether the large
$\mt$ approximation is adequate and whether Sudakov resummation is advantageous.
We argue that our results hold to all perturbative orders. Based on
it, we conclude that the full inclusion of finite top mass corrections
is likely to be important for accurate phenomenology for a light Higgs
with $\mh\sim125$~GeV at the LHC with $\sqrt s=14$~TeV.
\end{abstract}

\maketitle

An accurate computation of the Higgs boson production
cross-section~\cite{Dittmaier:2011ti} is essential in the search for
this particle, which might be on the verge of being discovered at the
LHC~\cite{higgssearch}.  If the Higgs is light, and in particular in
the region $\mh\sim125$~GeV, perhaps favored by LHC data, the dominant
Higgs production mechanism is gluon fusion, which starts at leading
order $\Ord(\as^2)$ through a (predominantly top) quark loop. Higher
order corrections, which turn out to be quite large, are accordingly
difficult to compute, and the full next-to-next-to-leading order
(NNLO) result is known either in the large $\mt$ limit, at the fully
differential level~\cite{Anastasiou:2005qj}, or as an expansion in
inverse powers of $\mt$ for the fully inclusive
cross-section~\cite{Harlander:2009mq}. As $\mt\to\infty$ the quark
loop shrinks to a point (pointlike approximation) and the LO process
becomes a tree-level process of an effective theory.

At NLO, where the exact result is known, the large $\mt$ approximation
turns out to work surprisingly well, even up to values of the Higgs
mass at and above the top pair production threshold. This result can
at least in part be understood based on the observation that the
partonic cross-section is dominated by logarithmically enhanced
contributions related to soft gluon radiation which are independent of
$\mt$, so the pointlike approximation becomes exact, up to an overall
factor which starts at NNLO~\cite{Kramer:1996iq}.  This soft dominance
should take place when $\mh$ is raised at fixed $s$ so the energy
$\hat s$ of the partonic production subprocess approaches threshold
$\hat s \sim \mh^2$.

Close enough to threshold it is advantageous to resum  these
logarithmically enhanced terms 
(threshold resummation), and it
has indeed been observed that this resummation significantly
corrects and stabilizes the perturbative result in regions in which
the pointlike approximation holds to satisfactory accuracy~\cite{Catani:2003zt}.
It is important to understand that this may happen even if the
expansion in powers of the strong coupling
$\as$ behaves in a perturbative way, i.e.\ if
the size of higher-order logarithmically enhanced contributions
decreases with the perturbative order, so an actual 
all-order resummation
is not really necessary. For this, it is sufficient that these enhanced
contributions approximate the missing higher orders well enough that their
inclusion actually improves the accuracy of the computation, and the
desirability of  resummation should thus be judged by the accuracy
of the logarithmic approximation.

Even so, this can only be possible if the center 
of mass energy of the partonic collision $\hat s$ is significantly
lower than the hadronic one $s$, which at the LHC is very far from
threshold. Because the gluon distribution is strongly peaked at small
values of the momentum fraction of each hadron carried by the
gluon itself, this is especially likely in a gluon fusion channel.  A
quantitative assessment of this effect is thus 
important  in order to determine the accuracy
of the pointlike limit, and also whether
threshold resummation is advantageous.

\begin{figure} 
\begin{center}
\includegraphics[width=\columnwidth]{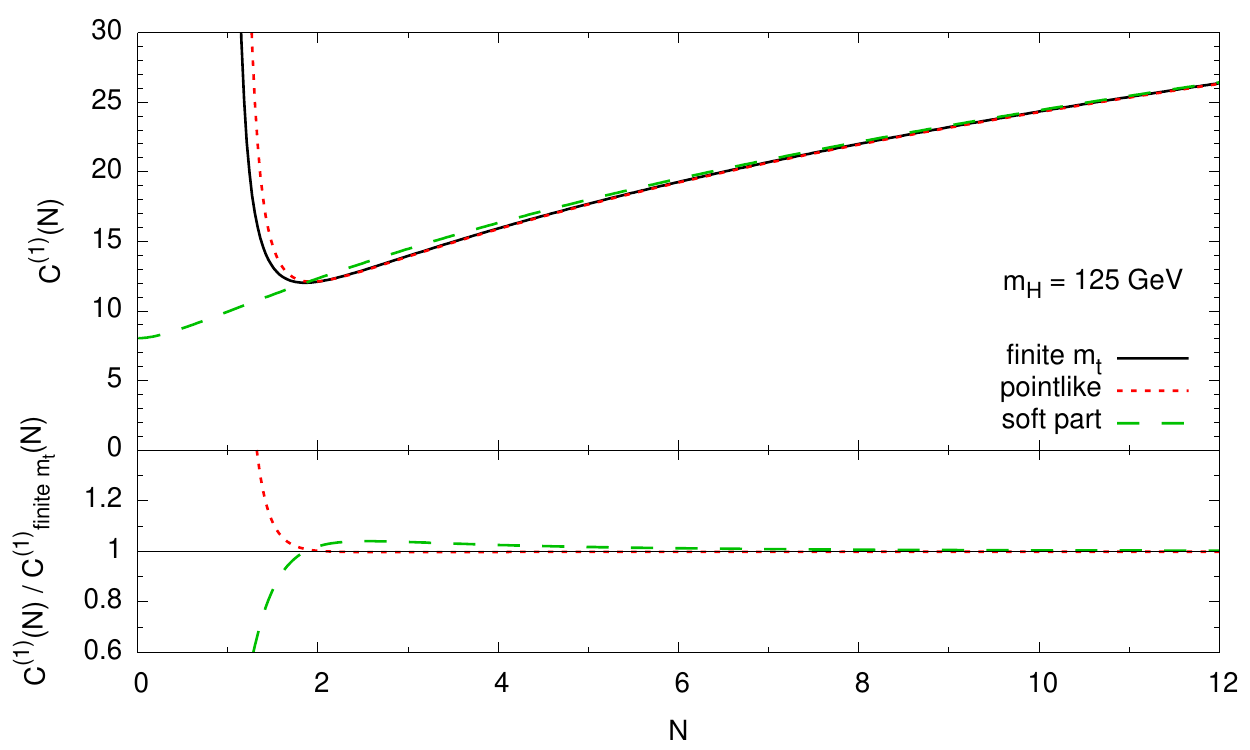}\\
\includegraphics[width=\columnwidth]{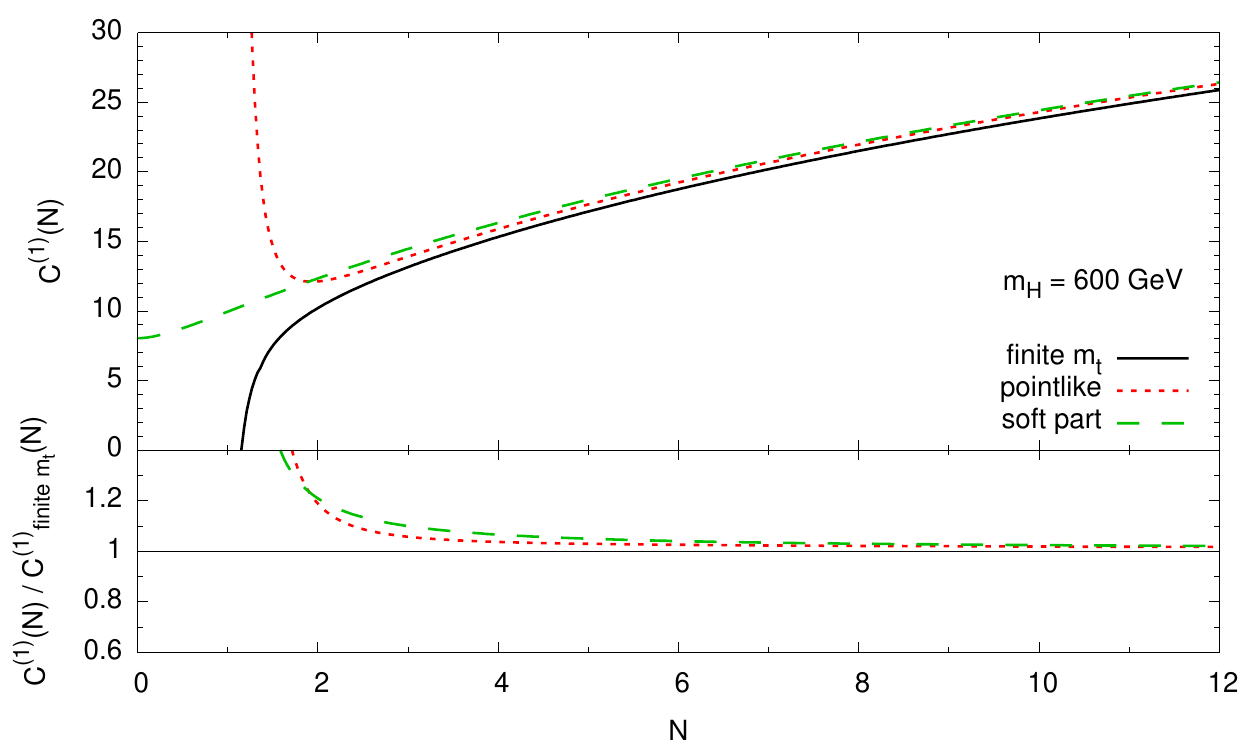}
\caption{The $\Ord(\as)$ contribution to the 
  coefficient function Eq.~\eqref{cofundef}
  as a function of $N$, for 
  $\mh=125$~GeV (upper plot) and $\mh=600$~GeV (lower plot).
  In each case we show the exact result and the pointlike and logarithmic
  approximations, as well as the ratio of the latter two to the
  exact result.}
\label{fig:Higgs_order_as}
\end{center}
\end{figure}

\begin{figure} 
\begin{center}
\includegraphics[width=\columnwidth]{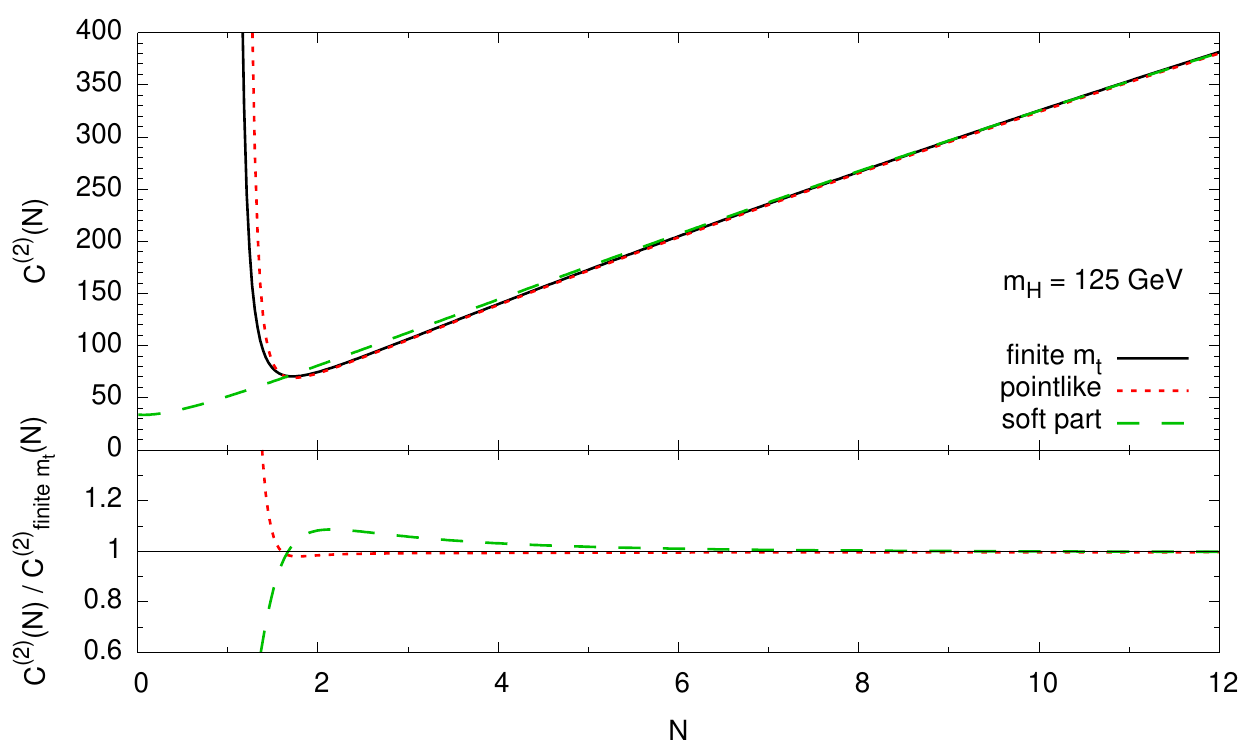}\\
\includegraphics[width=\columnwidth]{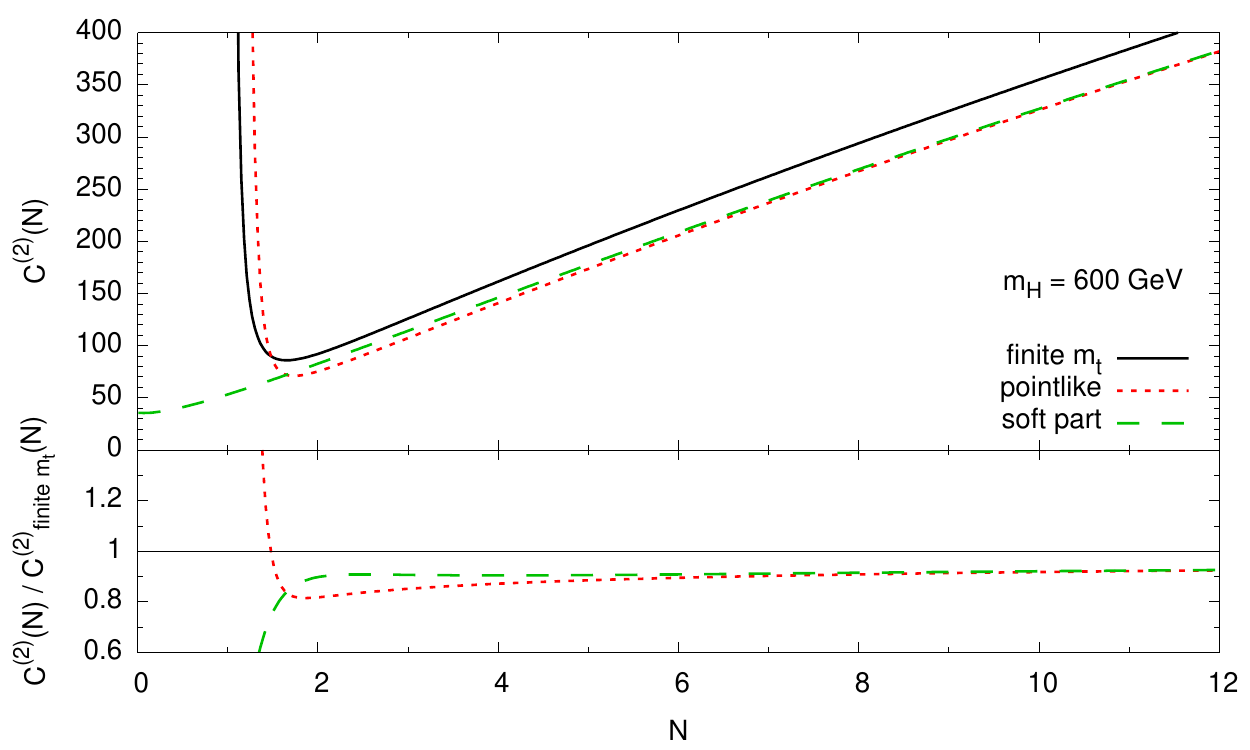}
\caption{Same as Fig.~\ref{fig:Higgs_order_as}, but to $\Ord(\as^2)$.
}
\label{fig:Higgs_order_as2}
\end{center}
\end{figure}

The necessary formalism was presented in Ref.~\cite{Bonvini:2010tp}, and
applied to Drell-Yan production.
The cross-section for Higgs production is a function of the scale
$m^2_H$ and a scaling variable $\tau=\mh^2/s$, given by  the 
convolution  
\beq
\label{eq:fact}
\frac{\sigma(\tau,\mh^2)}{\tau} = \int_\tau^1\frac{dz}{z} \,
\Lum\(\frac{\tau}{z},\mh^2\)\frac{\hat\sigma(z,\as(\mh^2))}{z}
\eeq
of a partonic cross-section $\hat\sigma\(\frac{\mh^2}{\hat s},\as(\mh^2)\)$
and a parton luminosity $\Lum\(\frac{\hat s}{s},\mh^2\)$;
a sum over relevant partonic subprocesses is understood.
In our case at LO the only subprocess is $gg\to H$, so
\beq
\label{eq:lumi}
\Lum(z,\mu^2)=\int_z^1\frac{dy}{y}\, g(y,\mu^2)\,
g\(\frac{z}{y},\mu^2\)
\eeq
where $g(z,\mu^2)$ is the gluon parton distribution (PDF) in the proton.
It is convenient to write the partonic cross-section 
in terms of a dimensionless coefficient function $C(z,\as)$
\beq
\hat\sigma(z,\as) = \sigma_0\, z\, C(z,\as),
\eeq
where  $\sigma_0$ is the LO partonic cross-section,
so  the $gg$ coefficient function has an expansion in powers of $\as$
\beq\label{cofundef}
C(z,\as) = \delta(1-z) + \as C^{(1)}(z) + \as^2 C^{(2)}(z) + \Ord(\as^3).
\eeq

Upon  Mellin transformation
\begin{align}
\sigma(N,\mh^2) &= \int_0^1d\tau\,\tau^{N-1}\,\frac{\sigma(\tau,\mh^2)}{\tau},\\
C(N,\as) &= \int_0^1dz\,z^{N-1}\,C(z,\as),\\
\Lum(N,\mu^2) &= \int_0^1dz\,z^{N-1}\,\Lum(z,\mu^2)
\end{align}
the convolution in Eq.~\eqref{eq:fact} turns into an ordinary product:
\beq
\label{eq:factN}
\sigma(N,\mh^2)=\sigma_0\, \Lum(N,\mh^2)\,C(N,\as).
\eeq
The Mellin transformation maps the large $\tau\to1$ region into the
large $N\to\infty$ region, and the small $\tau\to0$ region into the small
$N\to N_s$ region, with $N_s$ the rightmost singularity of $\sigma(N,\mh^2)$
(i.e.\ the convergence abscissa of the Mellin transform).
For gluon-initiated processes, $N_s=1$  to all perturbative orders.

The dominant partonic kinematic region can then be determined through a
saddle point argument, by computing the value of $N$ which
provides the  dominant contribution to the
Mellin inversion integral:
\beq\label{eq:imt}
\frac{\sigma(\tau,\mh^2)}{\tau} = \frac{1}{2\pi i} \int_{c-i\infty}^{c+i\infty} dN \,
\tau^{-N}\,\sigma(N,\mh^2) \qquad (c>N_s).
\eeq
Namely, we define
\beq\label{eq:melexp}
E(N,\tau,\mh^2) \equiv N\ln\frac{1}{\tau}+\ln \sigma(N,\mh^2),
\eeq
so that
\beq
\frac{\sigma(\tau,\mh^2)}{\tau} = \frac{1}{2\pi i} \int_{c-i\infty}^{c+i\infty} dN\,e^{E(N,\tau,\mh^2)}.
\label{eq:imt2}
\eeq
In the saddle point approximation, the 
integral 
Eq.~\eqref{eq:imt2} is dominated by the value
 of the exponent $E(N,\tau,\mh^2)$ at its stationary point $N_0$
\beq
\label{eq:saddle-point_def}
\left.\frac{\partial E(N,\tau,\mh^2)}{\partial N}\right|_{N=N_0}=0, 
\eeq
and by the behaviour in its vicinity.
The position of the saddle point is a function
$N_0=N_0(\tau,\mh^2)$, solution of
Eq.~\eqref{eq:saddle-point_def}. Hence, for any value of the physical
kinematics, the questions whether the partonic cross-section is 
well approximated by its pointlike limit or resummation is
advantageous are
answered by verifying whether this is the case for $C(N_0,\as)$.

A unique real saddle point is present due to the drop of the
cross-section $\sigma(N,\mh^2)$ as $N$ grows. This drop is
driven by the parton luminosity $\Lum(N,\mu^2)$, which thus controls
the position of the saddle $N_0$.  The drop of the
luminosity at large $N$ (and its growth at small $N$) reflects in turn
the drop of the PDFs and luminosity
as $z\to1$ (and their growth as $z\to0$).  However, for
large $N$, $C(N,\as)$ actually grows with $N$.  This growth,
which is due to the logarithmically enhanced contributions, is only
possible because the partonic cross-section is a distribution, rather
than an ordinary function: the Mellin transform of an
ordinary positive function is easily proven to be a decreasing
function of $N$.  Nevertheless, the parton luminosity always offsets this
increase, because the physical
cross-section $\sigma(\tau,\mh^2)$ is an ordinary positive function
and thus must decrease with $N$.  
The faster the growth of the cross-section, the better the saddle point approximation,
which is thus
especially good for gluon-dominated processes, as the gluon is
more peaked at small $z$ than the quark, and thus the gluon-gluon
luminosity drops faster than quark luminosity.

The stationary point
  is determined by the interplay in
Eq.~(\ref{eq:melexp}) of the
rise of the first term and the drop of the hadronic cross-section
$\sigma(N,\mh^2)$, which in turn is the product 
of the coefficient function and the luminosity.
The shape of the NLO and NNLO contributions to
the coefficient function
is shown in Figs.~\ref{fig:Higgs_order_as} and
\ref{fig:Higgs_order_as2},
for two values of the Higgs mass
which are allowed by present 
data~\cite{higgssearch}. Clearly, when the partonic cross-section
rises monotonically with $N$, 
it is the drop of the parton luminosity which determines 
the drop of the hadronic cross-section and thus
the position of the saddle, and even when it drops, the decrease of
its hadronic counterpart
is much stronger in the presence of a luminosity, so the
location of the saddle is substantially larger.

\begin{figure} 
\begin{center}
\includegraphics[width=\columnwidth,page=1]{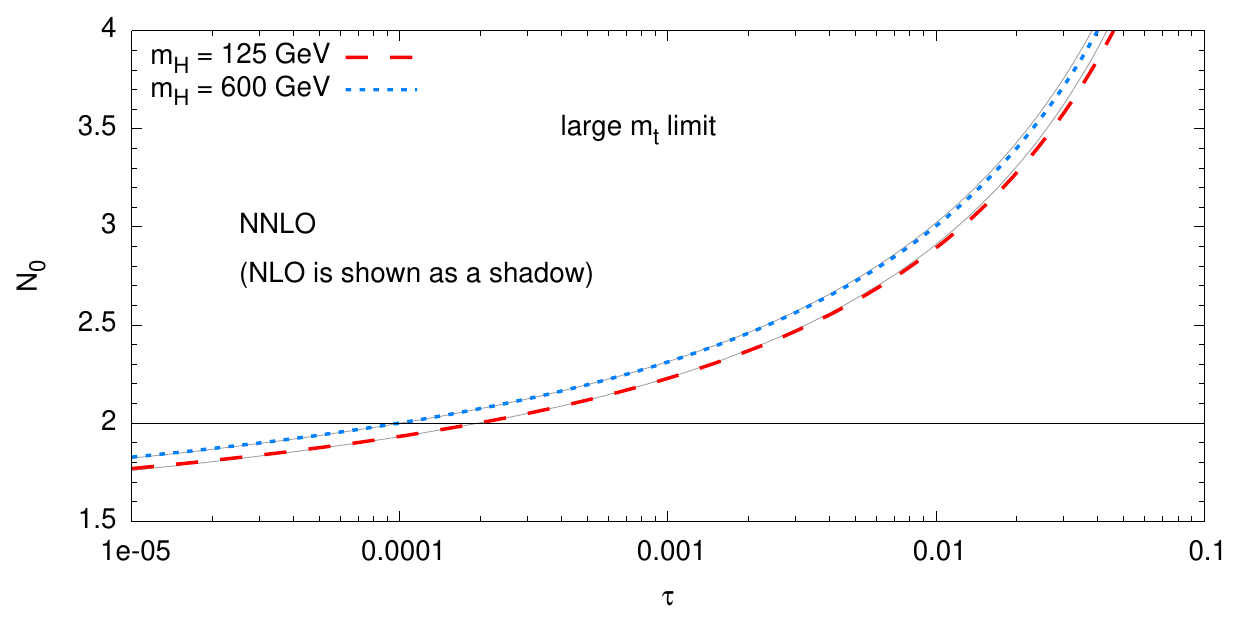}\\
\includegraphics[width=\columnwidth,page=3]{prl_Higgs_saddle-point}\\
\includegraphics[width=\columnwidth,page=2]{prl_Higgs_saddle-point}
\caption{
  The position of the saddle point $N_0$ for the Mellin inversion
  integral Eq.~\eqref{eq:imt} as a function of: 
  $\tau=\mh^2/s$ (top); $\mh$ for three different
  values of the collider energy $s$ (middle); $s$
  for two different values of $\mh$ (bottom). The curves in the top plot depend
  very weakly on either $\mh$ (shown) or $s$.
  Both the NLO and NNLO are shown, computed using the appropriate
  NNPDF2.1 PDF set~\cite{NNPDF21} in each case.}
\label{fig:sp}
\end{center}
\end{figure}

We have determined the position of the saddle at NLO and NNLO, using
NNPDF2.1 parton distributions~\cite{NNPDF21} at the corresponding order.
Results are shown in Fig.~\ref{fig:sp}, which is the main result of
this paper. 
The position of the saddle
$N_0$ depends on two independent variables, which can be chosen
out of the three variables $\mh$, $s$, and 
$\tau=\mh^2/s$.
If results for $N_0$ are shown as a function of
$\tau$, the dependence on the other variable (be it $\mh$ or $s$)
becomes very slight, because it enters only through the scale
dependence of $\as$ and the parton distributions: this is
explicitly seen in the top plot of Fig.~\ref{fig:sp}, where results
are shown for two different values of $\mh$; the dependence would be
similar if $\sqrt{s}$ were varied instead by a comparable factor. 
The dependence of the position of the saddle on the
perturbative order is completely negligible (as also  shown), and so is
the dependence on whether the pointlike approximation is used or
not. We have also checked that varying the
renormalization and factorization scales by a factor two changes the
value of $N_0$ by 1\% or less, even though more dramatic scheme
changes which affect the infrared behaviour of the PDF 
such as suggested in Ref.~\cite{deOliveira:2012qa} might have somewhat larger
effects. The impact of changing the PDF set or the value of
$\as$ in a reasonable range is rather less than that.

The size of the region around the value of $N_0$ which dominates the
integral can be estimated by computing the second derivative of
$\ln\sigma(N,\mh^2)$, which gives the width of the gaussian integral which
approximates the Mellin inversion in the
complex $N$ plane. We find that a one-sigma  region corresponds to
a variation of $N_0$ by about 25\% about its central
value. So, within the accuracy of our saddle point approximation, the
distinction between different curves at fixed $\tau$  is of little
import. For clarity and completeness, in Fig.~\ref{fig:sp}
we show the position of the saddle as a function of the Higgs
mass for the three values of $s$ relevant for the LHC, and as a
function of the center of mass energy for two values of the Higgs mass.

We can now  assess both the adequacy of the pointlike approximation,
and the desirability of
resummation for given values of $s$ and $\mh$: first, using
Fig.~\ref{fig:sp} the given hadronic kinematics can be translated
into a value of $N_0$. Then, for this  value
of $N_0$ we can check whether  
the pointlike approximation is accurate, and threshold
resummation is advisable. To this purpose, in 
Figs.~\ref{fig:Higgs_order_as}--\ref{fig:Higgs_order_as2} we compare
the exact NLO and NNLO contributions to the coefficient function to
its various approximations. At NLO, we use the
implementation of Ref.~\cite{Bonciani:2007ex} of the original exact
result of Refs.~\cite{Graudenz:1992pv,Spira:1995rr}; at NNLO a full
exact computation is not available, so for light Higgs we use the
expansion of Ref.~\cite{Harlander:2009mq} matched to the exact small
$z$ limit of Ref.~\cite{Marzani:2008az}, while for heavy Higgs, where
the expansion of Ref.~\cite{Harlander:2009mq} is unstable, the
``finite $\mt$'' curve of Fig.~\ref{fig:Higgs_order_as2} is merely
given by correcting the pointlike result through the inclusion of 
the first order correction in $\frac{\mt}{\mh}$
from  Ref.~\cite{Harlander:2009mq} (and it thus only indicates
the location of the region  where the
finite $\mt$ corrections are likely relevant). 

The pointlike results at NLO~\cite{Dawson:1990zj,Djouadi:1991tka} and
NNLO~\cite{Anastasiou:2002yz} have been long known. As a soft part, we
show all contributions which survive the $N\to\infty$ limit, defined
as the exact Mellin transform of all contributions of the form
$z\left(\frac{\ln^k\frac{(1-z)}{\sqrt{z}}}{1-z}\right)_+$ and all
contributions proportional to $\delta(1-z)$ to the NLO and NNLO
coefficient functions. While the coefficients of these terms
are fixed by soft resummation, there is a certain
latitude in defining which subleading terms to include. Our choice
reproduces the exact soft kinematics~\cite{Forte:2002ni,Bonvini:2010tp}
thereby optimizing the agreement with the exact result to all
orders. The region in which the soft limit defined in this way
is close to the full result is the region in which one expects soft
resummation to improve the accuracy of the computation,  
even when  all-order
resummation is not mandatory.

Inspection of
Figs.~\ref{fig:Higgs_order_as}--\ref{fig:Higgs_order_as2} shows that
the region in which the pointlike approximation is adequate is very
close to the region in which the soft approximation is good, so
indeed the success of the former might be explained by the
accuracy of the latter. Both at NLO and NNLO the relevant region is
roughly $N\gtrsim2$.  This is the region where the partonic 
cross-section starts to rise with $N$, driven by logarithmically
enhanced contributions.  The apparent failure of both approximations
for heavy Higgs at NNLO is likely to be due to the fact that in this
case the ``finite $\mt$'' computation is in fact only approximate. 

The
fact that the same behaviour is observed at NLO and NNLO is not
accidental. On the one hand, the position of the
saddle is largely determined by the parton luminosity, which is
perturbatively very stable at NLO and beyond~\cite{NNPDF21}. On the
other hand, the shape of the coefficient functions and the dominance
of soft terms are mostly controlled by the location of the leading
small- and large-$N$ singularities, which can be checked to be
stable even upon all-order resummation~\cite{Bonvini:2010tp}. This
supports the expectation that the desirability of resummation and the
all-order reliability of the pointlike approximation can be assessed
on the basis of the known low orders.

At the LHC, $\tau$ is quite small: if $\mh\sim125$~GeV,
$\tau\sim10^{-4}$ and if $\mh\sim600$~GeV,
$\tau\sim10^{-3}-10^{-2}$, thereby leading to values of $N_0$, close to
the transition value $N\sim2$, for which resummation is at best
desirable, but certainly not mandatory, as $\as\ln^2N_0\ll1$.
Nevertheless, Fig.~\ref{fig:sp}, in which the $N=2$ line has been
drawn for ease of reference, shows that for a heavy Higgs with
$\mh\sim600$~GeV the pointlike approximation is adequate, and Sudakov
resummation clearly advantageous, 
for any collider with center of mass
energy up to $\sqrt s \lesssim~30$~TeV, and thus certainly at the
LHC. 

Even for a light Higgs with $\mh\sim125$~GeV the pointlike
approximation is fine for a collider with energy of 7~TeV, but it
may start failing as the energy is raised, and becomes inadequate
at the LHC with $\sqrt s=14$~TeV, where the saddle drops at
$N_0\approx 1.9$
so the finite-mass corrections to $C^{(1)}$
Fig.~\ref{fig:Higgs_order_as} 
reach the percent level, though care should be taken because, close to
the region where the approximation breaks down, the conclusion
may depend on various details and approximations. However, because
the breakdown of the pointlike approximation is in significant part due
to the presence of spurious high-energy double
logs~\cite{Marzani:2008az}, 
it is expected to be
more noticeable in less inclusive observables. 
  Correspondingly, with this higher
center of mass energy, Sudakov resummation is likely to stop being
advantageous for a light Higgs. 

In summary, we have provided a way of assessing whether the pointlike
approximation is adequate and whether Sudakov resummation is
desirable based on the behaviour of the known first few orders of the
partonic cross-section. Based on it, we conclude that for a
light Higgs with $\mh\sim125$~GeV 
the full inclusion of finite top mass corrections to its
cross-section is likely to be important for accurate
phenomenology at the LHC with $\sqrt s=14$~TeV.
\begin{acknowledgments}
We thank F.~Tackmann for a critical reading of the manuscript, and
R.~Harlander and A.~Vicini for providing computing code.
\end{acknowledgments}

\end{document}